\documentclass[sigchi]{acmart}

\usepackage{booktabs} 
\usepackage{listings}
 
\setcopyright{none}

\renewcommand\footnotetextcopyrightpermission[1]{}

\begin{document}
\lstset{
	basicstyle=\ttfamily,
    breaklines=true,
    breakatwhitespace=true,
}

\title{Code Shrew}
\subtitle{Software platform for teaching programming through drawings and animations}

\author{Ludwik Trammer}
\affiliation{%
  \institution{Georgia Institute of Technology}
}
\email{ludwik@gatech.edu}

\author{Jamie Nunez}
\affiliation{%
  \institution{Georgia Institute of Technology}
}
\email{jamie@shrew.app}


\begin{abstract}
In this paper, we present Code Shrew, a new software platform accompanied by an interactive programming course. Its aim is to teach the fundamentals of computer programming by enabling users to create their own drawings and animations. The programming language has a straightforward syntax based on Python, with additions that enable easy drawing and animating using object-oriented code. The editor reacts seamlessly and instantly, providing an engaging and interactive environment for experimenting and testing ideas. The programming course consists of lessons that cover essential programming principles, as well as challenges to test users' skills as they progress through the course. Both the lessons and challenges take advantage of the editor's instant feedback, allowing for a focus on learning-by-doing. We describe the software and the content, the motivation behind them, and their connection to constructionism.
\end{abstract}

\maketitle

\section{Introduction}
We've built a web-based platform for learning and practicing the fundamental concepts of computer programming. \textit{Code Shrew} consists of those two main components:

\begin{itemize}
\item A programming environment which enables people to create drawings and animations with computer code. It uses a simple Python-like syntax and is designed to teach modern object-oriented programing techniques: one draws by creating objects and animates by changing their properties. Because of this characteristic, the resulting drawings and animations can also be seen as intuitive visualizations of the program's logic itself. They are displayed in real-time in the preview area.  Users can save them to their profile on the website and share them with friends on social media.

\item A course using the platform to present a set of fundamental programming concepts. Each lesson is based around a single central idea which is presented in a video, expanded in a text form, and then reinforced by interactive examples, which students can play with freely.
\end{itemize}

Among other things, the project was inspired by constructionism, a learning theory according to which people learn best by creating things \cite{papert1991situating}. That influenced both the choice to build a creative tool and the structure of the course - we wanted to give students the necessary tools to realize their own creative ideas, instead of forcing them to follow predefined paths.

\section{Related Work}
There are many existing solutions in this space. We discuss some of them here.

\subsection{Code Draw}
Georgia Tech's cs6460 project from Summer 2017 created by Michael Delfino \cite{delfino}. It was the initial inspiration behind our work. Delfino created a straightforward domain-specific language called Drawlang, intended for drawing shapes. The other part of his project was a multiuser web-based game inspired by Pictionary, called Code Draw. Players would use Drawlang code to draw pictures representing words, while other players would try to guess those words.

\subsection{Scratch}
A popular environment for teaching programming using draggable code blocks. Its value has been proven in numerous studies (including \cite{aivaloglou2016kids}, \cite{maloney2008programming}, \cite{saez2016visual}). One can use it to create games and animations by dragging sprints into the \textit{stage} area and then dragging code blocks to create programs. It's easy to use, but the programs created with it can be surprisingly sophisticated. Its website includes a lot of social features and can easily be seen as not only a programming environment, but also a social network based around creating, sharing and remixing visual programs.

\subsection{Alice}
Alice \cite{cooper2000alice}, similarly to MIT Scratch, is a block-based programming environment. It is however much more powerful and complex. Just like with Scratch, one creates Alice programs by dragging objects into a \textit{stage} area and then programming them with draggable code blocks. Unlike Scratch, Alice is object-oriented and aligned quite closely with a ``real'' programing language - Java. This closeness to Java can be viewed as both Alice's strength and weakness. On the one hand, the understanding gained from working with Alice can be easily transferred to ``real world'' programming with Java. On the other hand, this complexity makes Alice much more intimidating. Java code is known for having a very rigid complex structure, which can also be observed in Alice.

\subsection{Logo}
The original constructionist programming tool. One of its primary creators was Seymour Papert, the funder of the constructionist movement in education. For the 50 years since its creation, it was used to teach countless children all around the world and its effects had been studied extensively (including \cite{clements1987longitudinal}, \cite{noss1985creating}, \cite{miller1988effects}).

Logo programs are procedural - they consist of instructions that direct the behavior of a marker (usually depicted as a turtle) which travels across the canvas creating drawings along its path. This method can be used to create surprisingly attractive line drawings.

\subsection{Summary}
Each of those tools has its unique set of strengths and weaknesses. Our intention while building Code Shrew was not to replace them but to provide an alternative that focuses on those four areas:

\begin{enumerate}
\item Fostering creativity by making use of the constructionist methodology. We want students to develop their own personal, unique creations. They shouldn't be forced to follow predesigned paths.
\item A low barrier to entry. Simple things should be achievable with simple code, without needless boilerplate. An immediate visual feedback should be always provided.
\item Providing a good starting point for learning ``real'' programming tools later. That's why code used by Code Shrew is text-based, shares basic syntax with Python and is object-oriented.
\item Community. Students should be able to learn from each other by sharing and remixing their creations.
\end{enumerate}

\section{The software}
The software is publicly available under https://shrew.app. It works directly in the browser\footnote{Current versions of all major modern browsers are supported, including Google Chrome, Mozilla Firefox, Microsoft Edge, Apple Safari and Opera.} and does not require installation. Registration is not required to access most significant features, except for saving creations and marking creations as \textit{loved}.

The software has a responsive interface - it can be used on computers, tablets, and mobile phones. The interface automatically adapts to the screen size. When using the editor, a device with a physical keyboard is recommended. 

\subsection{The editor}
The editor is the place where drawings and animations are created. It features a large code editing area and a smaller preview area. The editor supports features known from other code editors such as code coloring, line numbering, and automatic indentation. The previews are generated continuously in real time so that user can always see an up-to-date reflection of their work. In case of an error in user's code, the offending line is marked with an icon and a description of the error. 

Drawings and animations created in the editor can be saved by clicking \textit{Save to my profile}. Saving requires a registered account (the drawings/animation itself can be made without an account, which can be created later as a step in the saving process). After the initial save, all the subsequent changes are preserved automatically.

The editor can be accessed by clicking the \textit{Create new} button in the top-right corner of the website or by clicking \textit{View code and remix} on the page of saved drawing/animation (or the \textit{Edit} button, when viewing one's own work). It can also be embedded on other pages, most notably in the course lessons and the documentation.

\subsection{The Language}
The language used to create drawings and animations is based on Python 3 \cite{van2011python}, which is a popular language used widely in the industry. The syntax for all the basic operations (including variable assignment, looping, conditional statements, and working with data structures) is shared with Python. Because of that, we will not discuss it in detail in this document. A fuller explanation is available in the \textit{Lessons} section on the website, under \href{https://shrew.app/lessons}{shrew.app/lessons}. Here we will focus on the drawing and animations parts of the language, that are unique to Code Shrew. That includes objects representing individual shapes, a context processor used with animations, and a separate module that can be utilized to import any of the free icons from \textit{FontAwesome.com}.  

\subsubsection{Shapes}
Shapes are the most basic building blocks of any Code Shrew program. Drawing is achieved by creating shape objects. For example, typing \texttt{Circle(color='red')} in the Code Shrew editor, will result in a red circle being displayed.

There are currently seven basic shapes in Code Shrew: \textit{Rectangle}, \textit{Square}, \textit{Ellipse}, \textit{Circle}, \textit{Line}, \textit{Polygon}, and \textit{Text}. The icons module (described later in this document) contains 766 more.

The appearance of a shape can be modified using its properties. Different kinds of shapes have different properties. A value of a property can be set when creating a shape or anytime later. Those are both equally valid methods of creating a brown square:

\textit{Method 1:}
\begin{lstlisting}
Square(color='brown')
\end{lstlisting}

\textit{Method 2:}
\begin{lstlisting}
john = Square()
john.color = 'brown'
\end{lstlisting}

Shapes also have methods that perform different actions. For example to create a mirror image of a text one can use the \texttt{flip\_horizontal()} method:

\begin{lstlisting}
hello = Text("Hello visitor!")
hello.flip_horizontal()
\end{lstlisting}

Shapes are drawn on a 100x100 plane. Of course, one can add as many shapes to a single drawing as they like.

Detailed descriptions of each shape (together with their individual properties and methods) can be found in the documentation under \href{https://shrew.app/documentation/shapes/}{shrew.app/documentation/shapes}.

\subsubsection{Icons}
Besides basic shapes, Code Shrew contains a powerful addition - the \texttt{icons} module. Icons behave just like shapes, but there are 766 of them, as the module enables the use of any of the free icons from the Font Awesome collection.

To use the module one has to import it first by writing \texttt{import icons}. That will cause Code Shrew to make the icons available. For example, the following code will draw a red track:

\begin{lstlisting}
import icons
icons.Truck(color='red')
\end{lstlisting}

More details on using the icons, together with their properties and methods can be found in the documentation under \href{https://shrew.app/documentation/icons/}{shrew.app/documentation/icons}.

\subsubsection{Colors}
Colors can be used as arguments for the \texttt{color} property on shapes and icons. Code Shrew supports solid color (i.e., one color) and gradients (i.e., multiple colors blending together).

Solid colors can be specified with text in quotes. The text should contain the color's name or its RGB value. That means that both methods shown below will result in a blue circle:

\textit{Method 1:}
\begin{lstlisting}
Circle(color='blue')
\end{lstlisting}

\textit{Method 2:}
\begin{lstlisting}
Circle(color='#0000FF')
\end{lstlisting}

In most cases, we recommend the first method as it is usually much easier to read and understand. Code Shrew know about 147 different color names, so referring to colors by name is not particularly limiting. The full list of available color names is available in the documentation under \linebreak \href{https://shrew.app/documentation/colors/}{shrew.app/documentation/colors}. As an example, the following code will create a fish colored \textit{salmon}:

\begin{lstlisting}
import icons
icons.Fish(color='salmon')
\end{lstlisting}

To create a gradient a list of solid colors is needed. For example, to create a circle gradually going from red to blue, the following code can be used:

\begin{lstlisting}
Circle(color=['red', 'blue'])
\end{lstlisting}

An arbitrary number of colors can be employed. For example, the following code will create a rainbow frog:

\begin{lstlisting}
import icons
icons.Frog(color=['red', 'orange', 'yellow', 'green', 'blue', 'purple'])
\end{lstlisting}

\subsubsection{Animations}
Creating animations is done by manipulating properties of shape and icon objects. The code needs to be inside the \texttt{with animation()} block, to indicate that the changes are to be animated.

For example, the following code will animate a circle by gradually changing its color from red to blue:

\begin{lstlisting}
my_circle = Circle(color='red')

with animation():
    my_circle.color = 'blue'
\end{lstlisting}

One can animate by changing almost any of the object's properties inside the \texttt{with animate()} block. That includes changing its color, size, position, rotation, transparency and other (the full list is available in the documentation).

By default, a \texttt{with animation()} block will take exactly 1 second to perform the animation. That can be changed by providing it a \texttt{duration}. For example, the following animation will take 5 seconds to run:

\begin{lstlisting}
my_circle = Circle(color='red')

with animation(duration=5):
    my_circle.color = 'blue'
\end{lstlisting}

Multiple properties can be animated at the same time. For example, this code will create a small rectangle going from left and change its color from pink to cyan, both at the same time:

\begin{lstlisting}
rec = Rectangle(width=20, height=20, x=10, color='deeppink')

with animation(duration=3):
    rec.x = 90
    rec.color = 'cyan'
\end{lstlisting}

There can be more than one \texttt{with animation()} block. They run one after another. For example, this code will make the rectangle go back and forth:

\begin{lstlisting}
rec = Rectangle(width=20, height=20, x=10, color='deeppink')

with animation(duration=2):
    rec.x = 90
    rec.color = 'cyan'
    
with animation(duration=1):
    rec.x = 10
    rec.color = 'deeppink'
\end{lstlisting}

Empty \texttt{with animation() blocks} are allowed. They will make the program wait for the duration of the block, without changing anything. Doing this requires adding the \texttt{pass} inside such block, to indicate that it is intended to be empty. For example, this will create an animation where a rectangle travels from left to right, waits for half a second and then goes back:

\begin{lstlisting}
rec = Rectangle(width=20, height=20, x=10, color='deeppink')

with animation(duration=2):
    rec.x = 90
    rec.color = 'cyan'
    
with animation(duration=0.5):
    pass
    
with animation(duration=1):
    rec.x = 10
    rec.color = 'deeppink'
\end{lstlisting}

Declarations of new shape objects are allowed inside the \texttt{with animation()} blocks and result in the shape appearing gradually for the duration of the block. For example, the following code will create an animation where it takes 5 seconds for the rectangle to appear fully:

\begin{lstlisting}
with animation(duration=5):
   rec = Rectangle(width=20, height=20, x=10, color='deeppink')

with animation(duration=2):
    rec.x = 90
    rec.color = 'cyan'
\end{lstlisting}

\subsection{Saved creation's page}
Each creation that has been saved has its own public page. The page includes:
\begin{itemize}
\item the drawing or animation itself
\item the date of creation
\item a link to the author's profile
\item if the creation is a remix: a link to the original creation and information about original creation's author
\item a \textit{See code and remix} button (or an \textit{Edit} button when viewed by the author) 
\item a \textit{Remove} button (only when viewed by the author)
\item a \textit{Share} button, which enables sharing to many different social media platforms
\item a \textit{Love} button with a counter (showing the number of people that ``loved'' the creation by clicking the button)
\end{itemize}

Many of those elements are centered around social interactions. We wanted people to be able to show off their creations (by sharing), appreciate creations of others and feel appreciated (by ``loving''), and get inspired by the work of others (by remixing). In a similar vein, we choose the saved creations that we like the most and feature them on the Code Shrew's homepage. This way we can show our appreciation to the authors, and inspire others by presenting what's possible. Even at the time of writing of this paper, right after the software was finished, we already got some very high-quality submissions from our users.

\subsection{User accounts}
While most of the website's functionality is available without registration, saving creations and marking creations as ``loved'' requires logging in. Users are able to log in using their existing Facebook or Google accounts or to create a new account by providing their email, the desired username, and password. 

Each user registered on the website has an individual public profile, listing their saved creations and creations that they ``loved''.

\subsection{Technological stack}
The technologies used in the project include (but are not limited to):

\begin{itemize}
\item \textit{Django (Python 3)} - the server-side framework,
\item \textit{PostgreSQL} - the database,
\item \textit{Bulma} - the CSS framework,
\item \textit{SASS} - the CSS extension language,
\item \textit{Babel} - the ES6 compiler,
\item \textit{Skulpt} - the in-browser Python interpreter,
\item \textit{svg.js} - the graphics library,
\item \textit{Code Mirror} - the code editor widget,
\item \textit{Cloud Flare} - the CDN and caching provider,
\item \textit{Markdown (CommonMark)} - the markup language,
\item \textit{CloudConvert} - the svg to png converter (used for  social media previews of creations).
\end{itemize}

\section{The course}
\subsection{Design}
Along with the Code Shrew tool and site, we created material for users to work through so they can learn what Code Shrew does as well as programming basics, providing a complete package within the Code Shrew website. Providing this side of the site will act as a quick way for new users to see how it all works and what areas they may be interested in trying out more. Also, examples of code are given in an interactive environment and explained in lesson videos, providing a useful resource for beginning programmers that are not sure where to begin.

Similar sites supply a programming tool alongside introductory programming lessons but these can be very limited due to cost (some require a membership or payment for full access to materials) and not employing complex or existing programming languages. In our case, we are offering this tool and lessons for free, using syntax similar to Python, and giving students a chance to learn about more complex structures such as for loops, functions, and objects. This will enable students to transition to scripting in Python (or similar languages) much more easily when they are ready to graduate from Code Shrew. This is a central idea behind Code Shrew and the lessons we developed: not only did we want to teach students basic programming rules in a very accessible and interactive way but also get them excited and ready for more complex programming tasks.

\subsection{Lessons}
The lessons created during this project cover how to use variables, lists, for loops, functions, and objects. Each of these are in respect to Code Shrew itself, so the user can experiment with the different programming techniques all within an engaging environment where the focus is on creating animations. Each lesson includes short videos that explain the new content at a high level, text that covers the same information, helpful links, examples, and a ``playground'' area where code is provided and the user can experiment with different values and rules. Users are able to change the code, see the result instantly and automatically, and even save their creation if desired.

The purpose of this lesson set up stems from the fact that, especially when it comes to programming, learning-by-doing is often a very successful and fun way to learn as it can be exciting to see what is possible and immediately implement any creative ideas that come to mind. Also, since our lessons provide different avenues of learning, students can tailor what components they interact with the most to how they learn best. For example, learning general content through the lesson video may be a great start for some students whereas trying some examples first, finding gaps in knowledge, and then checking out the lesson text for more information may work best for others. Overall, we wanted to facilitate learning in any way a student would feel most comfortable with since we both love programming and just want to give students a chance to see why it can be such an exciting field without getting intimidated.

\subsection{Challenges}
Programmers are often given tasks to complete, often with no exact guide of how it should be done. For example, they may be asked to create a software with certain capabilities or find an interesting story in an abundance of data. This is often, in our opinion, the funnest part of programming as it is a time to be creative, apply what you already know, and look into topics that aren't as familiar. With Code Shrew, we wanted to give users this same experience.

A Challenges tab is provided as an optional resource. Within this tab, the user will find designs and animations, but unlike the rest of the site, these do not come with the code needed to create them. They are then tasked with recreating the art to the best of their ability. To help with this, helpful (but minimal) parameters are provided to get them started. For example, exact shape sizes, colors of objects, and locations of certain pieces of the design/animation can be provided if they are not obvious. Also, provided with each of these is the level of skill needed to complete the task, provided as what lessons the user should already be familiar with and understand.

\subsection{What's Next?}
As a final component to the course content, there is a "What's Next?" resource where users can find information and links to other programming languages and tutorials. This part of the course is meant to direct the student when they are ready to move onto writing more challenging code, as well as show them that is indeed the next step, and there are many resources out there to help on them on this journey.

\section{Survey}
We created a survey and posted it on PeerSurvey, which is a virtual survey platform available to people with Georgia Tech accounts. The survey began with the following guidance:

\begin{quote}
To take this survey, you have to first visit \linebreak https://shrew.app and create an animation using the tool. It would be best if the code had at least 4 lines but, beyond that, it may be very creative or very simple - it's your choice. To save the animation you will need to log in. You may use your existing Google or Facebook account, or create a new account on the site.
\end{quote}

It then presented those 11 questions:

\begin{enumerate}
\item Paste the URL of the animation that you created on shrew.app and saved to your profile
\item When learning how to create the animation, have you visited the ``Lessons'' section?
\item (if yes) Where the lessons helpful? Do you have any suggestions?
\item When learning how to create the animation, have you visited the "Documentation" section?
\item (if yes) Where the documentation helpful? Do you have any suggestions?
\item When learning how to create the animation, have you looked at the featured creations on the homepage?
\item (if yes) Where they helpful? Do you have any suggestions?
\item What is your age?
\item What is your level as a programmer?
\item When working with the tool and the site, have you experienced any problems?
\item Do you have other suggestions or ideas?
\end{enumerate}

It is important to note that the results are not representative. The survey was meant as quick User Experience testing, not a scientific research. We received 31 responses, which gave us interesting insights, including bug reports and suggestions for improvements. We were able to implement most of them already.

While we asked people to create an animation, we did not suggest a way to learn how to do it. That was on purpose - we wanted to see what will people do organically. From all the people surveyed, 66\% indicated that they used the \textit{Lessons} section, 69\% used the documentation, and 56\% used examples featured on the homepage. All people who answered the question asking if their chosen method of learning was ``helpful'' indicated that it was. We did not get a single ``not helpful'' answer. It is hard to tell if this is a reflection of the quality of our work, or just of Georgia Tech community being inherently supportive.

\section{Limitations}
A limitation of the course content created was that it was not created by someone who was not trained to teach, meaning some items could have been presented in a much better way to help facilitate learning. We remedied this as best we could by discussing the lessons and how they should be approached as well as getting feedback through the survey discussed here. Additionally, Ludwik was a teacher in the past so, while he did not create the videos, he was able to provide feedback on different ideas for general lesson set up and course design.

Another big limitation of the project as a whole is the fact that there was no time to test it with users. A proper study is needed to explore if what we created is compelling and if it has positive effects on programming skills. 

\section{Future Work}
\subsection{The course}
This course was created over the span of a single semester and there is still so much more that can be done to improve it with more time. For example, additional lessons could be added that cover while loops, booleans, conditional statements, how to read documentation, if/else statements, working with data, and more. A fun environment like Code Shrew really provides a lot of space for creative lessons and fun examples. As a potential remedy, we feel this a perfect area for someone to add to this project in the future and hope a future student, or team, can contribute more content to give Code Shrew a more comprehensive, well-rounded course.

\subsection{The software}
There are many possible directions for further development of the website. Some of them include:

\begin{enumerate}
\item Supporting more advanced drawing features (for example grouping, controlling outlines,  and clipping). Adding anything that is supported by the SVG standard should be pretty straightforward from the technical standpoint. The biggest challenge is doing it while preserving the simplicity of the language.
\item Social features that would allow for more communication between the users. For example comments section under saved creations. Such communication can be very beneficial in a constructionist education \cite{bruckman1998community}, but keeping the environment safe and friendly will be a challenge.
\item Adding support for sound so that animations can also include sound effects.
\item Allowing users to provide new challenges for our \textit{Challenges} section.
\item Support typing suggestions (\textit{autocomplete}) in the editor. We already have a working version of this feature, but it is not able to provide adequate suggestions in some contexts, so it is disabled for now (we do not want to confuse users who might start to rely on this future). To achieve this, we have built a custom code parser built on top of the \textit{nearley.js} framework. A more advanced parser is needed.
\item Provide the code created by users with information about the environment (e.g., mouse movements, keyboard and mouse events). That would allow people to create interactive games using our platform.
\end{enumerate}

\bibliographystyle{ACM-Reference-Format}
\bibliography{bibliography}

\end{document}